\documentclass{webofc}
\usepackage[varg]{txfonts}   
\usepackage{bm} \usepackage{bbold} \usepackage{multirow} \usepackage{diagbox}
\usepackage{lineno} \usepackage[normalem]{ulem} 
\usepackage{mathtools}
\begin{document}
\title{QCD static force in gradient flow}
%
%

\author{\firstname{Xiang-Peng} \lastname{Wang}\inst{1}\fnsep\thanks{\email{xiangpeng.wang@tum.de}}
}

\institute{Physik Department, 
Technische Universität München,
         James-Franck-Str. 1,  85748 Garching, Germany
          }

\abstract{%
We review our recent study on the QCD static force using gradient flow at next-to-leading order in the strong coupling.  The QCD static force has the advantage of being free of the $O(\Lambda_{\text{QCD}})$ renormalon appearing in the static potential but suffers from poor convergence in the lattice QCD computations. It is expected that the gradient flow formalism can improve the convergence. Based on our next-to-leading-order calculations, we explore the properties of the static force for arbitrary flow time $t$, as well as  in the limit $t\rightarrow 0$, which may be useful for lattice QCD simulations. 
}
\maketitle
\section{Motivation}
\label{motivation}
The QCD static potential encodes important information about the QCD interactions for wide range of distances~\cite{Wilson:1974sk, Susskind:1976pi, Fischler:1977yf, Brown:1979ya}. 
At short distances, the static potential can be calculated perturbatively~\cite{Fischler:1977yf,Schroder:1998vy,Brambilla:1999qa,Anzai:2009tm,Smirnov:2009fh} and has been proved to be useful in extracting the strong coupling constant $\alpha_s$ from the latticee QCD simulations~\cite{Karbstein:2014bsa,Bazavov:2014soa,Karbstein:2018mzo,Takaura:2018vcy,Bazavov:2019qoo,Ayala:2020odx}. The perturbative calculation of the static potential in dimensional regularization suffers a renormalon of order $\Lambda_{\rm QCD}$, which may
be absorbed in an overall constant shift~\cite{Pineda:1998id, Hoang:1998nz}. Analogously, in lattice regularization, there is a linear divergence that is
proportional to the inverse of the lattice spacing.  

The static force defined by the spatial derivative of the static potential has the advantage of being free from order $\Lambda_{\rm QCD}$ rernormalon, which make it more convenient for comparing lattice simulations with perturbative calculations~\cite{Necco:2001xg,Necco:2001gh,Pineda:2002se}. The force can be computed from the finite differences of the lattice data of the static potential, which works well if the available data are dense, like in
the case of quenched lattice data~\cite{Necco:2001xg}. However, at short distances, the lattice data are still sparse, and the
computation of the force from their finite differences leads to large uncertainties~\cite{Bazavov:2014soa}. An other way of calculating the force is to 
compute the force directly from a Wilson loop with a chromoelectric field insertion in it~\cite{Vairo:2015vgb, Vairo:2016pxb}.
However, direct lattice QCD calculations of the static force exhibit sizable discretization
errors and the convergence to the continuum limit is rather slow~\cite{Brambilla:2021wqs}. This poor
convergence may be understood from the convergence of the Fourier transform of
the perturbative QCD calculation in momentum space at tree level, in which the spatial momentum integration $\int_{0}^{C} d \bm{q} \cos(\bm{q} r) = \frac{\sin(C r)}{r}$ does not converge to a fixed value in the limit $C\rightarrow +\infty$, which indicates the the cut off regularization in lattice simulations will face convergence problem in the continuum limit.

The gradient-flow formulation has been proved useful in lattice QCD calculations of
correlation functions and local operator matrix
elements~\cite{Narayanan:2006rf, Luscher:2009eq, Luscher:2010iy,Luscher:2011bx,
Luscher:2013vga, Borsanyi:2012zs, Suzuki:2013gza,Makino:2014taa,
Harlander:2018zpi}. In gradient flow, the gauge fields in the operator
definitions of matrix elements are replaced by flowed fields that
depend on the spacetime coordinate and the flow time $t$. At tree level in
perturbation theory, the flowed fields come with a factor $e^{- q^2 t}$ for
every momentum-space gauge field with momentum $q$ in the operator definition, which gives an exponential suppression factor at large $\bm{q}$ in the Fourier transformation. 
If this supprersion behavior is kept unspoiled beyond tree
level, we expect that the poor convergence of the lattice QCD calculation of
the static force will be greatly improved by using gradient flow.

In this proceeding, we compute the static force in gradient flow in perturbation
theory at next-to-leading order in the strong coupling.  This calculation is
significant in two aspects.  First, we examine the convergence of the Fourier
transform explicitly beyond tree level.  Second, we examine the dependence on
the flow time $t$, and in particular the behavior in the limit $t \to 0$, that
may be useful when extrapolating to QCD from lattice
calculations done in gradient flow~\cite{Leino:2021vop}. 


\section{Definitions and conventions}
\label{sec:defs}
We define the QCD static potential in Euclidean space as~\cite{Wilson:1974sk,
Susskind:1976pi, Fischler:1977yf, Brown:1979ya} 
\begin{equation} V(r) = -
\lim_{T \to \infty} \frac{1}{T} \log \langle W_{r \times T} \rangle,
\end{equation} 
where $W_{r \times T}$ is a Wilson loop with temporal and
spatial extension $T$ and $r$, respectively, and $\langle \cdots \rangle$ is
the color-normalized time-ordered vacuum expectation value 
\begin{equation}
\langle \cdots \rangle = \frac{\langle 0 | {\cal T} \cdots | 0 \rangle}
{\langle 0 | {\rm tr}_{\rm color} {\bf 1}_{\rm c} | 0 \rangle}, 
\end{equation} 
with ${\cal T}$ the time ordering, $|0\rangle$ the QCD vacuum, and ${\rm tr}_{\rm
color} {\bf 1}_{\rm c}  = N_c$  the number of colors.  An explicit
expression for the Wilson loop $W_{r \times T}$ is 
\begin{equation}
\label{eq:wloop} W_{r \times T} = {\rm tr}_{\rm color} P \exp \left[ i g
\oint_C dz^\mu A_\mu (z) \right], 
\end{equation} 
where $P$ stands for the path
ordering of the color matrices, $A_\mu$ is the bare gluon field, $g$ is the
bare strong coupling, and $C$ is a closed contour.


We  define the QCD static
force by the spatial derivative of $V(r)$ as ~\cite{Brambilla:2000gk,Vairo:2015vgb,
Vairo:2016pxb,Brambilla:2019zqc,Brambilla:2021wqs}
\begin{equation}
\label{eq:forcedef} 
F(r) \equiv \frac{\partial}{\partial r} V(r) = -i \lim_{T
\to \infty} \frac{1}{T} \frac{\int_{-T/2}^{+T/2} dx_0 \,\langle W_{r \times T}
\hat{\boldmath{r}} \cdot g \boldmath{E}(x_0, \boldmath{r}) \rangle}{\langle W_{r \times T}
\rangle}, 
\end{equation} 
where in the second equality, the chromoelectric field
$g E_i$ is inserted into the Wilson loop at spacetime point
$(x_0,\boldmath{r})$.  

The static force in gradient flow, $F(r;t)$, can be defined by replacing the
gluon fields $g A_\mu(x)$ by the flowed fields $B_\mu(x;t)$, where $B_\mu$ is
defined through the flow equation~\cite{Luscher:2010iy, Luscher:2011bx}
\begin{eqnarray}
\frac{\partial}{\partial t} B_\mu (x;t) = D_\nu G_{\nu \mu}
+ \lambda D_\mu \partial_\nu B_\nu,\, G_{\mu \nu} = \partial_\mu B_\nu -
\partial_\nu B_\mu + [B_\mu, B_\nu], \,  D_\mu = \partial_\mu + [B_\mu,
\cdot], 
\end{eqnarray} 
with the initial condition \begin{equation} B_\mu
(x;t=0) = g A_\mu(x).  \end{equation} Here, $\lambda$ is an arbitrary constant,
and the flow time $t$ is a variable of mass dimension $-2$. 

For perturbative calculations of the static force, it is advantageous to first compute the static
potential in gradient flow, $V(r;t)$, by replacing the gluon fields $g A_\mu(x)$ by the flowed fields $B_\mu(x;t)$ in the definition of the Wilson
loop and choose $\lambda=1$, and then differentiate with respect to $r$ to obtain $F(r;t)$. Since the contributions from the spatial-direction Wilson lines at the times $\pm T/2$
vanish in the limit $T\rightarrow \infty$ in Feynman gauge~\cite{Schroder:1999sg},  we will employ the Feynman gauge in the calculation
of the static potential. The momentum-space potential
$\tilde{V}(\bm{q})$ is related to the position-space counterpart by
\begin{equation} 
V(r) = \int \frac{d^3\bm{q}}{(2 \pi)^3} \tilde{V}(\bm{q}) e^{i\bm{q} \cdot \bm{r}},
 \end{equation} 
 which is also valid for the static potential in gradient flow.  We neglect contributions to $\tilde{V}(\bm{q})$
with support only at vanishing spatial momentum $\bm{q}$, such as the heavy quark/antiquark self-energy diagrams, because they do not contribute to the static force. 

It has been shown that the IR divergences cancel in the sum of all Feynman diagrams up to two loops~\cite{Brambilla:1999qa,Brambilla:1999xf}, 
and the UV divergences can be removed by the renormalization of the strong coupling. We adopt dimensional regularization with spacetime dimension $d=4-2\epsilon$ and renormalize the strong coupling in the $\overline{\rm MS}$ scheme.

\section{Next-to-leading order results}
\label{sec:diags}

\begin{figure}[ht] 
\begin{center}
\includegraphics[width=0.9\textwidth]{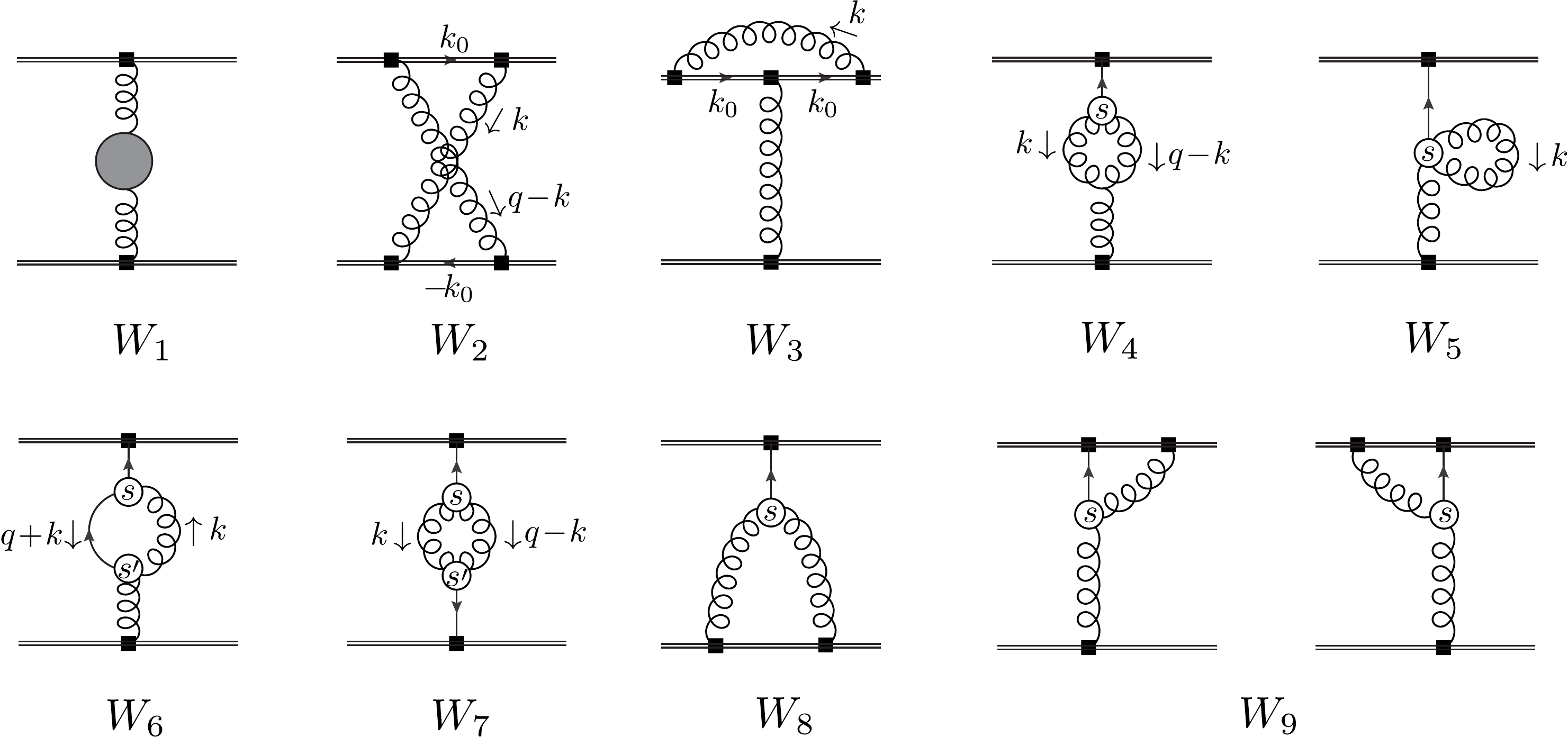}
\caption{\label{fig:potential_flow} Feynman diagrams for the static potential
at next-to-leading order in $\alpha_s$.  The double lines represent temporal
Wilson lines, curly lines are gluons, and solid lines are flow lines. Filled
squares represent the flowed $B_\mu$ fields at flow time $t$, and open circles
are flow vertices.  The blob in $W_1$ represents the gluon vacuum
polarization.  } 
\end{center} 
\end{figure}

In Feynman gauge, the Feynman diagrams for the static potential at next-to-leading order (NLO) in $\alpha_s$ are shown in Figure~\ref{fig:potential_flow}.  
After renormalization of the strong coupling in the $\overline{\rm MS}$ scheme,  up to order $\alpha_s^2$, we obtain
 \begin{eqnarray}
\tilde{V}(\bm{q};t) = - \frac{4 \pi \alpha_s (\mu) C_F e^{-2 \bm{q}^2 t}
}{\bm{q}^2} \bigg\{ 1+ \frac{\alpha_s (\mu)}{4 \pi}
\bigg[ \beta_0 \log(\mu^2 /\bm{q}^2) + a_1 + C_A \, W_{\rm NLO}^F (\bar{t})
\bigg]\bigg\} , 
\end{eqnarray}
where $\bar{t} \equiv \bm{q}^2 t$,  $\beta_0 = \frac{11}{3}C_A - \frac{2}{3} n_f$, 
$a_1 =  \frac{31}{9}C_A - \frac{10}{9}n_f$ and $W_{\rm NLO}^F (\bar{t})$ is finite function of $\bar{t}$. We are not able to obtain full analytical expression for $W_{\rm NLO}^F (\bar{t})$. 
\begin{figure}[ht] 
\begin{center}
\includegraphics[width=0.6\textwidth]{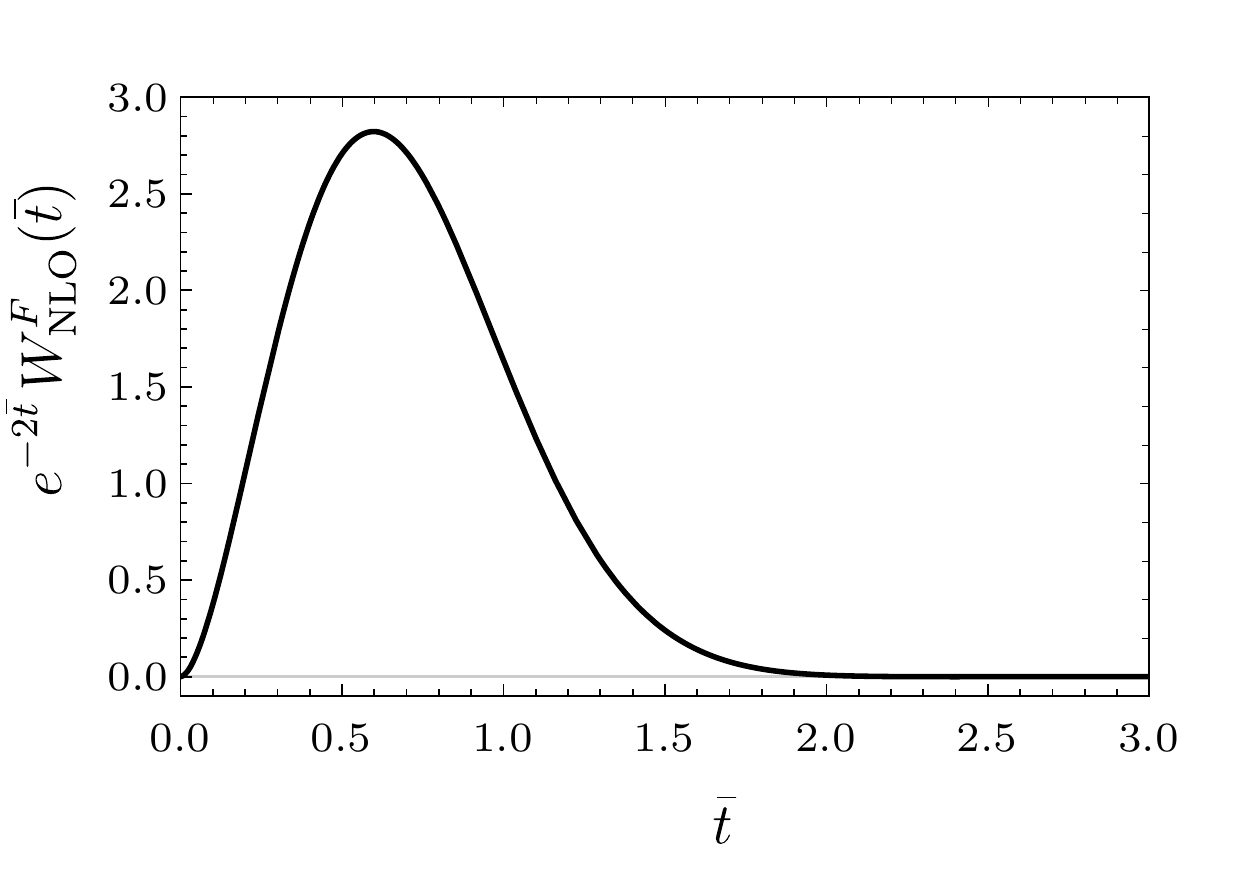}
\caption{\label{fig:WFNLOplot} The finite correction term $e^{-2 \bar{t}}
W_{\rm NLO}^F (\bar{t})$ as a function of $\bar{t} = \bm{q}^2t$.  }
\end{center} 
\end{figure}
Here we only show the numerical results for $ e^{-2 \bar{t}} \, W_{\rm
NLO}^F (\bar{t})$ in Figure~\ref{fig:WFNLOplot}. (See ref. \cite{Brambilla:2021egm} for integration representation of $W_{\rm NLO}^F (\bar{t})$.) 

It is interesting to see the behavior of $W_{\rm NLO}^F (\bar{t})$ in the limit $\bar{t}\rightarrow 0$. And we obtain
\begin{equation}
\label{eq:WNLOF_exp} W_{\rm NLO}^F (\bar{t}) = \bar{t} \left( -\frac{22
\gamma_{\rm E} }{3}+\frac{277}{18}-\frac{31 \log 2}{3} - \frac{22}{3} \log
\bar{t} \right) + O(\bar{t}^2).
\end{equation} 

The static force in position space is expressed as
 \begin{equation} 
 F(r;t) =
\frac{\partial}{\partial r} \int \frac{d^3\bm{q}}{(2 \pi)^3}
\tilde{V}(\bm{q};t) e^{i \bm{q} \cdot \bm{r}} = \frac{1}{r^2} \int_0^\infty d
|\bm{q}| \bm{q}^2 \frac{|\bm{q}| r \cos(|\bm{q}| r) - \sin(|\bm{q}| r)}{2 \pi^2
|\bm{q}|} \tilde{V}(\bm{q};t).  
\end{equation} 
We define the following dimensionless quantities
\begin{eqnarray} 
{\cal F}_0 (r;t) &=& -\int_0^\infty d |\bm{q}| \bm{q}^2
\frac{|\bm{q}| r \cos(|\bm{q}| r) - \sin(|\bm{q}| r)}{2 \pi^2 |\bm{q}|} \frac{4
\pi e^{-2 \bm{q}^2 t}}{\bm{q}^2}, \\ {\cal F}_{\rm NLO}^L (r;t;\mu) &=&
-\int_0^\infty d |\bm{q}| \bm{q}^2 \frac{|\bm{q}| r \cos(|\bm{q}| r) -
\sin(|\bm{q}| r)}{2 \pi^2 |\bm{q}|} \frac{4 \pi e^{-2 \bm{q}^2 t}}{\bm{q}^2}
\log (\mu^2/\bm{q}^2), \\ {\cal F}_{\rm NLO}^F (r;t) &=& -\int_0^\infty d
|\bm{q}| \bm{q}^2 \frac{|\bm{q}| r \cos(|\bm{q}| r) - \sin(|\bm{q}| r)}{2 \pi^2
|\bm{q}|} \frac{4 \pi e^{-2 \bm{q}^2 t}}{\bm{q}^2} W_{\rm NLO}^F(\bar{t} =
\bm{q}^2 t), 
\end{eqnarray} 
so that, up to order-$\alpha_s^2$, 
\begin{eqnarray} F(r;t) =
\frac{\alpha_s(\mu) C_F}{r^2} \bigg[ \left( 1 + \frac{\alpha_s}{4 \pi} a_1
\right) {\cal F}_0 (r;t) + \frac{\alpha_s}{4 \pi}
\beta_0 {\cal F}_{\rm NLO}^L (r;t;\mu) + \frac{\alpha_s C_A}{4 \pi} {\cal
F}_{\rm NLO}^F (r;t) \bigg].  
\end{eqnarray} 
We obtain the behavior of $r^2F(r;t)$ near $t=0$ as 
\begin{equation} 
\label{eq:asym}
r^2 F(r; t) \approx r^2
F(r;t=0) + \frac{\alpha_s^2 C_F}{4 \pi} \left[ - 12 \beta_0 - 6 C_A c_L\right]
\frac{t}{r^2}, 
\end{equation} 
where $c_L = -22/3$ is the coefficient of
$\bar{t} \log \bar{t}$ in Eq.~\eqref{eq:WNLOF_exp} and
 $F(r;t=0)$ is the usual
 QCD order-$\alpha_s^2$ result for the static force 
\begin{equation} 
F(r; t=0) =
\frac{\alpha_s(\mu) C_F}{r^2} \left\{ 1+ \frac{\alpha_s}{4 \pi} \left[ a_1 +2
\beta_0 \log (\mu r e^{\gamma_{\rm E}-1}) \right]\right\}.
\end{equation} 
Surprisingly, the coefficient of the $t/r^2$ term is $\left[- 12 \beta_0 - 6 C_A c_L \right] = 8 n_f$, which vanishes in the pure SU(3) gauge theory ($n_f=0$).

\begin{figure}[ht] \begin{center}
\includegraphics[width=0.55\textwidth]{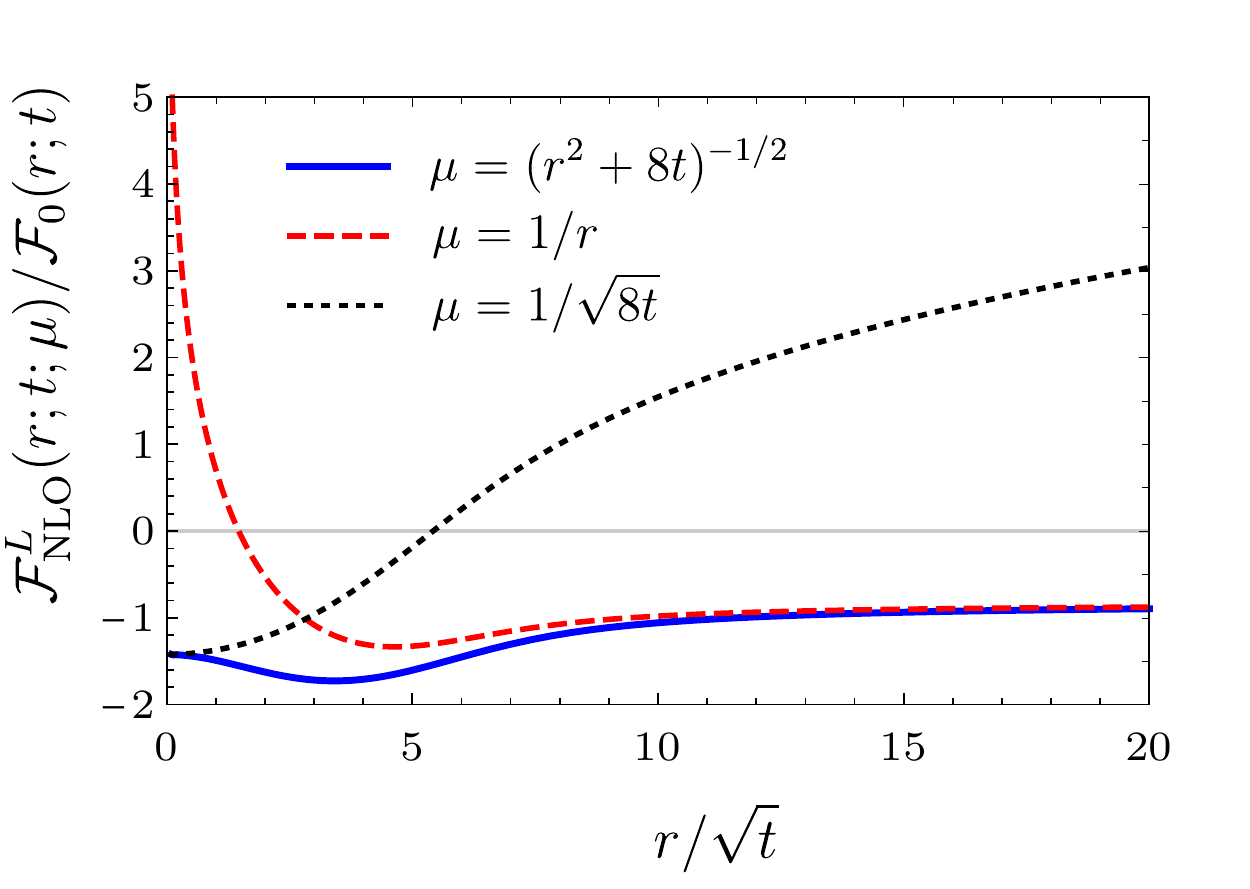} \caption{\label{fig:FLmudep}
The logarithmic correction factor ${\cal F}_{\rm NLO}^L (r;t;\mu)/{\cal
F}_0(r;t)$ for $\mu = ( r^2 + 8 t )^{-1/2}$, $\mu = 1/r$, and $\mu = 1/\sqrt{8
t}$ shown as a function of $r/\sqrt{t}$.  }
\end{center} 
\end{figure}

We show  ${\cal F}_{\rm NLO}^L (r;t;\mu)/{\cal F}_0(r;t)$ as a function of $r/\sqrt{t}$ in Figure~\ref{fig:FLmudep} with different choices of renormalization scales.
We find that the scale choice $\mu = ( r^2 + 8 t )^{-1/2}$ makes the logarithmic correction
factor ${\cal F}_{\rm NLO}^L (r;t;\mu)/{\cal F}_0(r;t)$ of order 1 for all
values of $r/\sqrt{t}$. Therefore, we will use $\mu = ( r^2 + 8 t )^{-1/2}$ here and below for numerical analysis.

\begin{figure}[ht] \begin{center}
\includegraphics[width=0.48\textwidth]{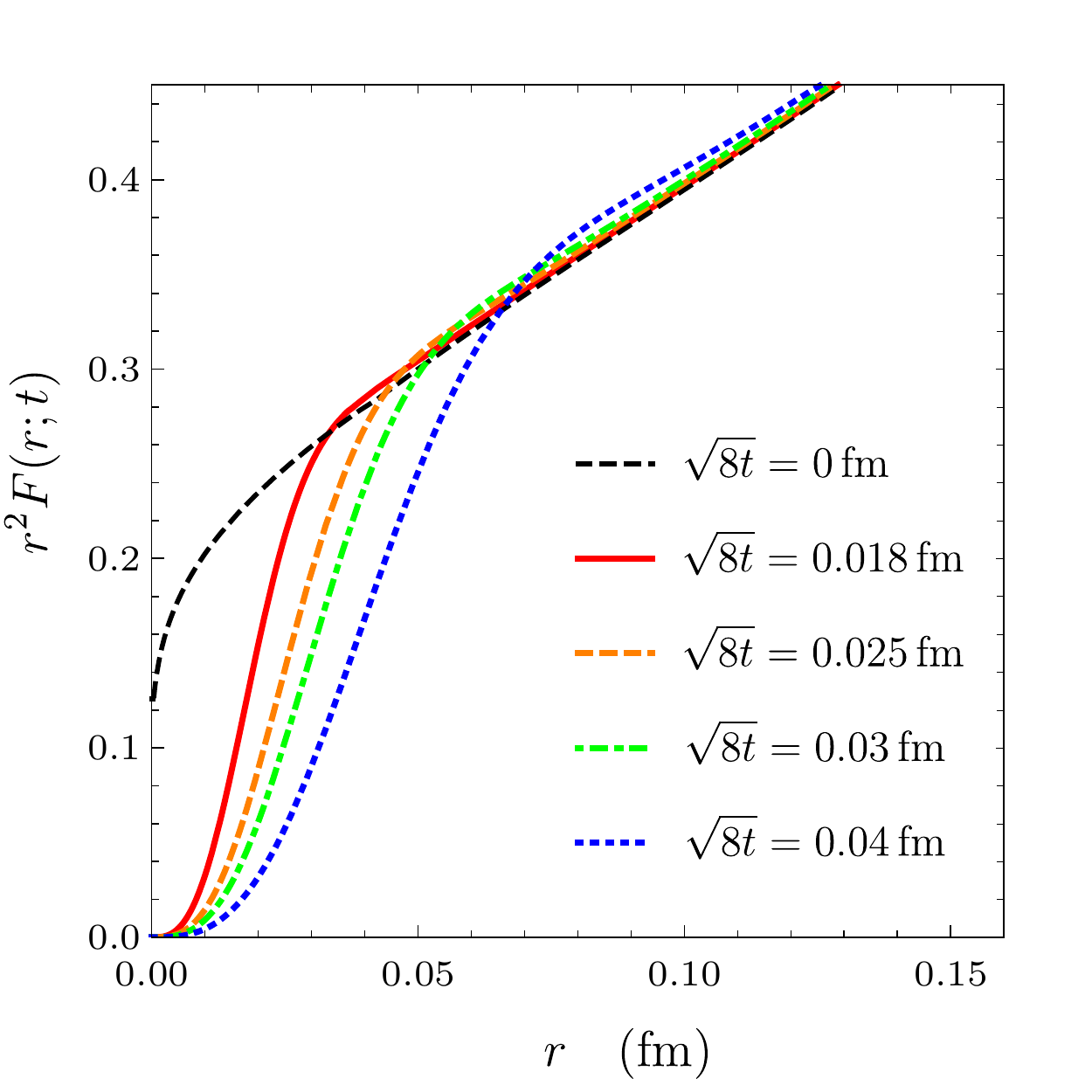}
\includegraphics[width=0.48\textwidth]{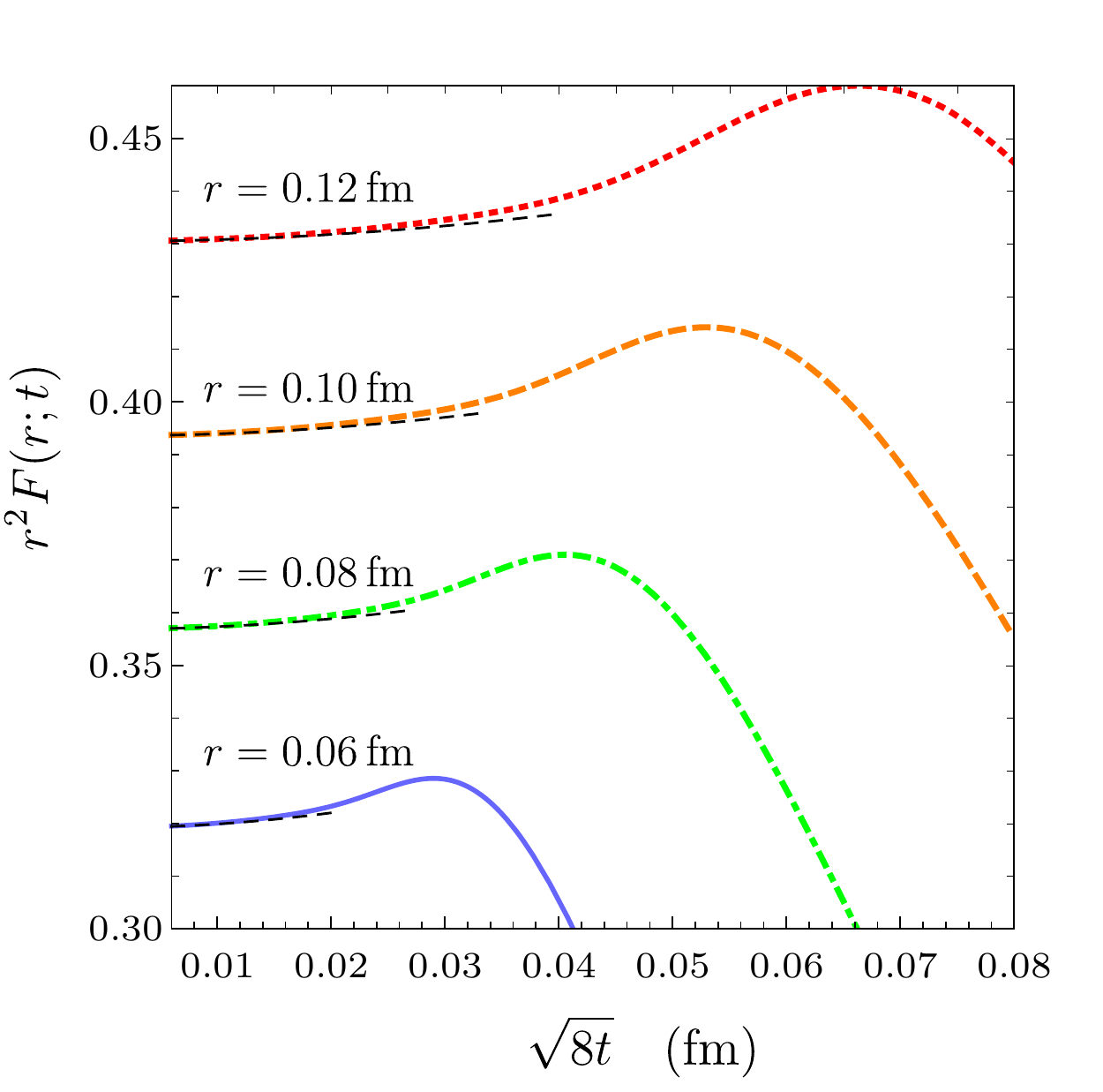}
\caption{\label{fig:rtdepplot} Left panel: numerical results for $r^2 F(r;t)$
for fixed values of $\sqrt{8 t}$ as functions of $r$.  Right panel: numerical
results for $r^2 F(r;t)$ for fixed values of $r$ as functions of $\sqrt{8 t}$;
the black dashed lines are approximate results based on Eq.~\eqref{eq:asym},
which is valid at small flow time.  We have set $\mu = (r^2+8 t)^{-1/2}$ and
$n_f = 4$.  } 
\end{center} 
\end{figure}

Computing $\alpha_s$ in the $\overline{\rm
MS}$ scheme at the scale $\mu = (r^2+8 t)^{-1/2}$ by using {\sf
RunDec}~\cite{Chetyrkin:2000yt} at four loops, and setting $n_f = 4$, we show the numerical results for the static force in gradient flow, $r^2
F(r;t)$, at NLO in $\alpha_s$ in Figure~\ref{fig:rtdepplot}.
In the left panel of Figure~\ref{fig:rtdepplot} , we show $r^2 F(r;t)$ as a
function of $r$ for several fixed values of $\sqrt{8 t}$.  We can see that $r^2
F(r;t)$ vanishes for $r \to 0$, while we recover the QCD result $r^2 F(r;t=0)$
(black dashed line) for $r \gg \sqrt{8 t}$. 
In the right panel of Figure~\ref{fig:rtdepplot}, we show $r^2 F(r;t)$ as a
function of $t$ for several fixed values of $r$.  As $t$ decreases, $r^2 F(r;t)$ approaches to the QCD result
as expected. We compare the exact NLO result for $r^2 F(r;t)$ at small $t$ with the
expression given in Eq.~\eqref{eq:asym} (black dashed lines) in the right panel of
Figure~\ref{fig:rtdepplot}. 
\begin{figure}[ht] \begin{center}
\includegraphics[width=0.45\textwidth]{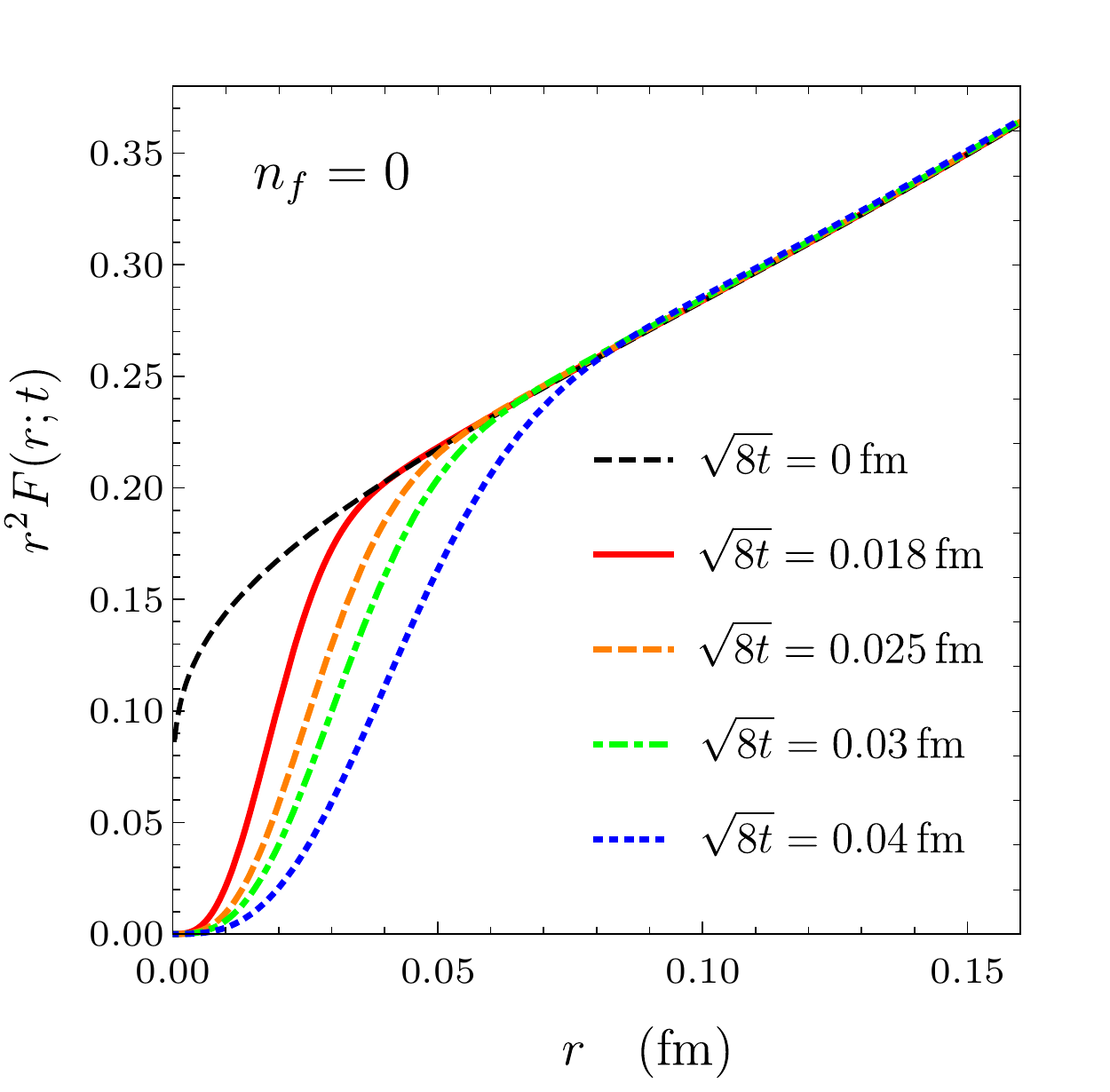}
\includegraphics[width=0.45\textwidth]{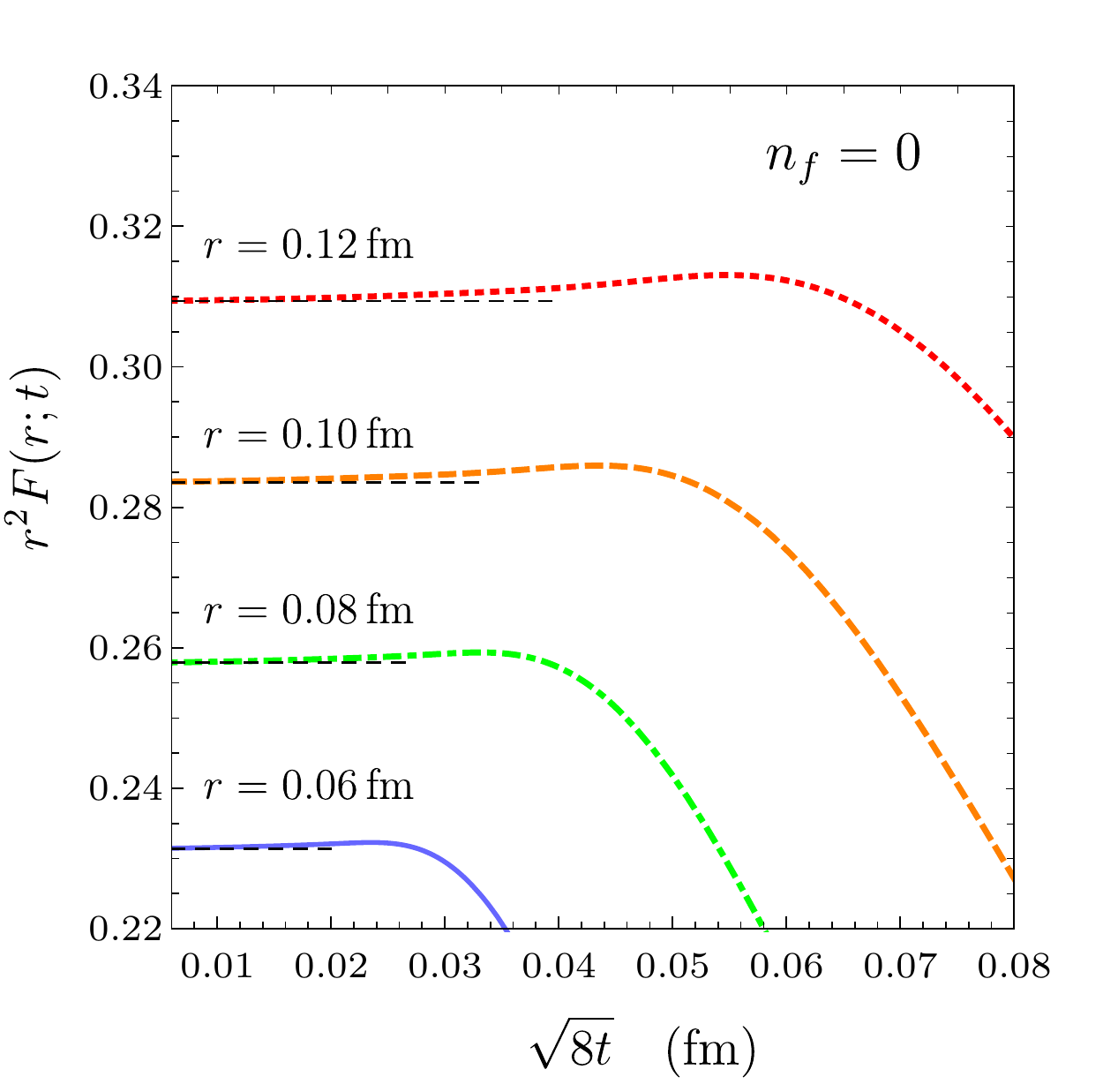}
\caption{\label{fig:rtdepplot_quenched} Left panel: numerical results for $r^2
F(r;t)$ in the pure SU(3) gauge theory ($n_f =0$) for fixed values of $\sqrt{8
t}$ as functions of $r$.  Right panel: numerical results for $r^2 F(r;t)$ in
the pure SU(3) gauge theory ($n_f =0$) for fixed values of $r$ as    functions
of $\sqrt{8 t}$; the black dashed lines are the QCD results for $r^2 F(r;t=0)$.
We have set $\mu = (r^2+8 t)^{-1/2}$.  } \end{center}
\end{figure}

We also show $r^2 F(r;t)$ in the pure SU(3) gauge theory ($n_f=0$, quenched case) in
Figure~\ref{fig:rtdepplot_quenched} as a function of $r$ for fixed values of
$\sqrt{8t}$, and as a function of $\sqrt{8 t}$ for fixed values of $r$.  We
have computed $\alpha_s$ in the pure SU(3) gauge theory by using {\sf
RunDec}~\cite{Chetyrkin:2000yt} at four loops, based on the value $r_0
\Lambda_{\rm QCD} = 0.637^{+0.032}_{-0.030}$ in ref.~\cite{Brambilla:2010pp},
with $r_0 = 0.5$~fm~\cite{Necco:2001xg}.  Due to the vanishing of the term
linear in $t$ in Eq.~\eqref{eq:asym}, the expression in Eq.~\eqref{eq:asym} is
equal to the QCD result $r^2 F(r;t=0)$, which we show in the right panel of
Figure~\ref{fig:rtdepplot_quenched} as horizontal black dashed lines.

\section{Summary}
\label{sec:summary}
In this proceeding, we review our recent study on QCD static force in gradient flow at NLO in the strong coupling. As we have anticipated in the motivation, the gradient flow indeed makes the Fourier transform of the static force in momentum space better converging. Thus we expect that the use of gradient flow  may also improve the convergence towards the continuum limit of the lattice QCD simulation of static force done at finite flow time, which is also supported by the results in
ref.~\cite{Leino:2021vop}. Our analytic results of the static force in the limit $t\rightarrow 0$ will also be useful when extrapolating to QCD from lattice calculations done in gradient flow. 
Similar analyses could be
extended to a vast range of nonperturbative quarkonium observables in the
 factorization framework provided by nonrelativistic effective field
theories~\cite{Brambilla:2004jw}.  For instance, one could study in gradient
flow the quarkonium potential at higher orders in
$1/m$~\cite{Eichten:1979pu,Barchielli:1988zp,Bali:1997am,Brambilla:2000gk,Pineda:2000sz,Brambilla:2003mu,Koma:2006si,Koma:2006fw,Koma:2010zza},
where $m$ is the heavy quark mass, static hybrid
potentials~\cite{Juge:2002br,Capitani:2018rox,Schlosser:2021wnr}, hybrid
potentials at higher orders in
$1/m$~\cite{Oncala:2017hop,Brambilla:2018pyn,Brambilla:2019jfi}, gluonic
correlators entering the expressions of quarkonium inclusive widths and cross
sections~\cite{Brambilla:2001xy,Brambilla:2002nu,Brambilla:2020xod,Brambilla:2020ojz,Brambilla:2021abf}.

\section*{\acknowledgementname}
I thank Nora Brambilla, Hee Sok Chung and Antonio Vairo for collaboration on the work presented here. The work of X.-P. W. is supported by the DFG (Deutsche Forschungs- gemeinschaft, German Research Foundation) Grant No. BR 4058/2-2 and by the  the DFG cluster of excellence “ORIGINS” under Germany’s Excellence Strategy - EXC-2094 - 390783311.
%
\bibliography{force}

\end{document}